\documentclass{JHEP3}
\usepackage{epsfig}

\newcommand{\be}{\begin{equation}}
\newcommand{\ee}{\end{equation}}
\newcommand{\bea}{\begin{eqnarray}}
\newcommand{\eea}{\end{eqnarray}}
\newcommand{\IR}{\mathbb{R}} 

\def\IZ{\relax\ifmmode\hbox{Z\kern-.4em Z}\else{Z\kern-.4em Z}\fi}
\newcommand{\IS}{{\bf S}}

\newcommand{\non}{\nonumber \\}

\def\half{{1 \over 2}} 

\def\del{{\partial}}

 \def\bh{{\bar h}}

\def\hA{\hat{A}} \def\hC{\hat{C}} \def\hL{\hat{L}}
\def\tg{\tilde{g}} \def\tR{\tilde{R}}
\def\wttriangle{\widetilde{\triangle}}

\def\al{\alpha} \def\bt{\beta}
  \def\eps{\epsilon}

\preprint{{\tt hep-th/0309190}}

\title{ \center{}Caged Black Holes:  \\ Black Holes in
Compactified Spacetimes I -- Theory}

\author{Barak Kol, Evgeny Sorkin, Tsvi Piran
\\
 Racah Institute of Physics\\
 Hebrew University \\
 Jerusalem 91904,
 Israel\\
{\tt barak\_kol, sorkin, tsvi @phys.huji.ac.il}}

\abstract{
In backgrounds with compact dimensions there may exist
several phases of black objects including the black-hole and the
black-string. The phase transition between them raises puzzles and
touches fundamental issues such as topology change, uniqueness and
Cosmic Censorship. No analytic solution is known for the black
hole, and moreover, one can expect approximate solutions only for
very small black holes, while the phase transition physics happens
when the black hole is large. Hence we turn to numerical
solutions. Here some theoretical background to the numerical
analysis is given, while the results will appear in a forthcoming
paper. Goals for a numerical analysis are set. The scalar charge
and tension along the compact dimension are defined and used as
improved order parameters which put both the black hole and the
black string at finite values on the phase diagram. Predictions
for small black holes are presented. The differential and the
integrated forms of the first law are derived, and the latter
(Smarr's formula) can be used to estimate the ``overall numerical
error''. Field asymptotics and expressions for physical quantities
in terms of the numerical ones are supplied. Techniques include
``method of equivalent charges'', free energy, dimensional
reduction, and analytic perturbation for small black holes.
}

\begin{document}




\section{Introduction}

In the presence of extra compact dimensions, there may exist
several phases of massive solutions of General Relativity,
depending on the relative size of the object and the relevant
length scales in the compact dimensions. For concreteness, we
consider a background with a single compact dimension --
$\IR^{d-2,1} \times \IS^1$. We denote the coordinate along the
compact dimension by $z$ and the period by $\hL$. The problem is
characterized by a single dimensionless parameter, for
instance\footnote{Later we will define an alternative parameter
$x$.} the dimensionless mass $\mu=G_N M/\hL^{d-3}$ where $G_N$ is
the $d$ dimensional Newton constant and $M$ is the mass (measured
at infinity).

The relevant phases are
\begin{itemize}
 \item The uniform black string.

A string solution is one which has an $S^{d-3} \times \IS^1$
horizon topology. The uniform string is $z$-independent and
 it is described by the $d-1$
Schwarzschild - Tangherlini metric \cite{Tangherlini} with the
addition of $+dz^2$ -- a spectating $z$ coordinate. This solution
was shown to be unstable to gravitational perturbation below a
certain critical $\mu$ by Gregory and Laflamme (GL)
\cite{GL1,GL2}.

 \item The non-uniform string.

These are solutions with the string horizon topology, which are no
longer $z$-independent. The marginally unstable GL - mode implies
that such a branch of solutions is emanating from the GL critical
point (for instance due to Morse theory \cite{TopChange}), and
{\it a priori} there could be other branches as well. Gubser
determined that the phase transition at the GL point is first
order by an analytic perturbation analysis \cite{Gubser}. Morse
theory implies that the emanating branch would be unstable, at
least at the beginning \cite{TopChange}. Finally these solutions
were found numerically by Wiseman in 6d \cite{Wiseman1},
who managed to formulate axially-symmetric gravitostatics (namely,
essentially 2d) in a ``relaxation'' form (a procedure familiar
from electrostatics) while presenting the constraints through
``Cauchy-Riemann - like" relations.

 \item The black hole.

These are solutions with $\IS^{d-2}$ horizon topology. For small
(dimensionless) mass we expect these solutions to resemble a $d$
dimensional black hole near the horizon. No analytic solution is
known, and we consider our numerical solutions in 5d
\cite{numerical} as strong evidence for their existence
(indications appeared already in \cite{PiranSorkin}).

 \item The stable non-uniform string.

Horowitz and Maeda \cite{HorowitzMaeda1} postulated the existence
of an additional stable phase, to serve as the end-point for the
GL decay.
This phase should be distinguished from the unstable non-uniform
string which is presumably a consequence of Gubser's analysis (or
from strings which are too massive to serve as the end-point for
decay -- clearly the stability of the solution is critical for the
physical interpretation).
Horowitz and Maeda argued, based on theorems for the increasing
area of the event horizon, that horizon pinching is impossible. By
now there is mounting evidence against the existence of this
phase: in \cite{TopChange} it was shown that a continuous topology
change (as a function of a parameter rather than time) is actually
possible, moreover Morse theory implies an obstacle, namely that
the addition of this phase must be accompanied by some other one,
in \cite{Wiseman1} it was 
 shown numerically that the whole branch of static non-uniform
strings originating from the GL point is too massive to serve as
an end-point, finally a full time-dependent numerical simulation
had to stop at $R_{\mbox{min}}/R_{\mbox{max}} \sim 1/13$ (due to
grid stretching)
without finding 
 clear
 evidence for stabilization towards this phase \cite{CLOPPV}
\footnote{Note however that the authors of this paper did not
interpret their results either as supporting or as countering the
conjecture.}
 ($R_{\mbox{max}},\, R_{\mbox{min}}$ refer to the
$z$-dependent radius of the horizon) . Additional circumstantial
evidence is supplied by Gubser's frustrated hope to find a second
order transition \cite{Gubser} and by the failed attempt to find
the non-uniform string analytically by looking for 5d
algebraically - special solutions \cite{deSmet}.

Our interpretation is that the argument of \cite{HorowitzMaeda1}
fails at a deep and interesting level -- the horizon theorems
being used there require the singularity not to leave the horizon,
while it seems to us plausible that this system constitutes an
example for  violation of Cosmic Censorship.

\end{itemize}

In addition to the research described above we would like to
mention: an ansatz that reduces the number of unknown functions
due to Harmark and Obers \cite{HO1}, speculations for a
generalization of uniqueness to $d>4$ \cite{uniqueness}, some
phenomenological implications for huge explosions accompanying the
transition \cite{explosive}, numerical evidence for the ideas of
\cite{TopChange} including the approach to a cone
\cite{Wiseman2,KolWiseman}, a time-symmetric initial data study
\cite{PiranSorkin}, a numerical study of black holes on a brane
which have a similar geometry \cite{KudohTanakaNakamura} a
recently discovered, and a related instability in highly rotating
black holes \cite{EmparanMyers}.
Other related research included a relation between the GL
transition and gauge theories through Matrix M-Theory
\cite{SusskindGW}, a prediction for a non-uniform near-extremal
black string \cite{HorowitzMaeda2}, a useful review of numerical
relativity \cite{Lehner_review}, a relation between the GL
instability and the negative heat capacity of black holes
\cite{Reall}, a discussion of implication of the GL instability to
more general spacetimes with horizons \cite{Hubeny:2002xn}, a
discussion of the explosion expected during these phase transition
\cite{Casadio:2000py}, a recent analysis of 4d metrics with a
compact dimension but different asymptotic conditions and a study
of non-uniform strings in the two-brane system
\cite{TamakiKannoSoda}. See also
\cite{Argyres:1998qn,Emparan:2000rs,Giddings:2001bu} for a short
and non-representative list of papers on black holes and large
extra dimensions/ brane worlds/ accelerator prospects.

Here, motivated by \cite{TopChange} we take the route of
approaching the phase transition by increasing the mass of the
black hole, rather than through the GL instability. Since no
analytic solution is known for the black hole we turn to numerical
solutions. Moreover, even though one can expect approximate
analytic solutions to exist for very small black holes, the phase
transition physics happens when the black hole is of the same size
as the compact dimension, and so a numerical simulation is
essential. It is important to list the goals of such a numerical
study, even if currently we cannot reach all of them

\begin{itemize}
 \item Establish the existence of the black hole solution.
 \item Determine the maximal (dimensionless) mass for a black hole
(it is expected not to fit in the compact circle beyond a certain
size \cite{TopChange}). Identify the direct reason for that.
 \item Establish another failure of generalized uniqueness, by
 demonstrating a black hole solution with the same mass as some
 (stable) black string.
 \item Test the description of approach to topology change proposed in
 \cite{TopChange}, namely that the geometry in the region between the two poles
approaches a cone and that at least for $5 \le d < 10$ the
geometry destabilizes (perturbatively) before reaching transition.
 \item To determine whether the black string decay violates Cosmic
Censorship is one of the most interesting issues pertaining to the
system. However, it is not clear if the study of black holes would
contribute directly.
 \end{itemize}

This paper contains theoretical background and results related to
the black hole solutions. After describing our ansatz for the
metric in section \ref{ansatz_section} we turn to an analysis of
the asymptotics and charges at infinity. We start in section
\ref{equivalent_sources_section} (inspired by
\cite{MyersPerry,MyersCompact}) with what we term ``a method of
equivalent sources'' where one pretends that gravity is weak
throughout and the source is large and smeared. We identify the
numerically measured asymptotics both in harmonic and in conformal
gauge and find their relation with the physical charges, the mass
and the tension. We promote a physical picture of the tension
being a tension of an imaginary string, which acts to decrease the
size of the extra dimension counteracting the effect of the mass
which ``wants to make space'' for itself. We also stress the
importance of the tension as a natural coordinate for the system's
phase diagram instead of the previously used $\lambda=\half
[(R_{\mbox{max}}/R_{\mbox{min}})-1]$) one that has the advantage
of putting both the black hole and the black string on the same
phase diagram, being always finite. This section ends by making
some predictions for small black holes.

In section \ref{free_energy_section} we use a rigorous technique
based on the (Gibbons-Hawking) free energy to revisit some of the
issues of the previous section and to derive the first law.
Together one arrives at a form of the first law, $dm=T\, dS  +
\tau\, d\hL$ which is completely analogous to the familiar $dE =
T\, dS -P \, dV$ (for gas) (see also \cite{Townsend:2001rg}). This
(differential) form of the first law is then readily combined with
scale invariance to yield the integrated form of the first law
(also known as ``Smarr's formula''). The latter is an important
test for numerics since it ties quantities at the horizon with
those at infinity and relies on the satisfaction of the equations
of motion everywhere (since the derivation uses integration by
parts) and so it gives a measure for the ``overall numerical
error''.

Next, in section \ref{dim_reduction_section} we use the fact that
in the asymptotic region the $z$ dependence is lost to perform a
dimensional reduction. The mass defined in section
\ref{equivalent_sources_section} is seen to agree with the
standard definition of mass in the lower dimension. The asymptotic
constant for the decay of the size of the extra dimension is
interpreted in the lower dimension as a scalar charge. The scalar
charge, like the tension is a natural parameter for the system's
phase diagram and it is especially useful for the Gregory-Laflamme
transition since it vanishes for the string and is non-zero for
all other known solutions. It shares another interesting property
with the tension: the tension, much like the mass, is always
positive \cite{PositiveTraschen,PositiveSIT} and it seems that so
is the scalar charge -- it would be interesting to prove/ refute
this.

In section \ref{at_horizon} we turn to the horizon, identify the
measurable quantities, and give predictions for small black holes.
We conclude by summarizing our main results in section
\ref{summary}. The appendices include details of a coordinate
transformation, yet another confirmation of the expression for the
mass (this time using the Hawking-Horowitz expression), the
details of action evaluation in the harmonic gauge and a different
proof of Smarr's formula.

Very recently the paper \cite{HO2} appeared which overlaps with some
of these theoretical considerations. Our 5d numerical results will
follow in a sequel paper \cite{numerical}.  Since the appearance of
the first version of this paper, another interesting paper appeared
\cite{HO3}.

\section{Ansatz for metric}
\label{ansatz_section}

The system under study, a static (no angular momentum) black
object in ${{\IR}}^{d-2,1} \times {\IS}^1$, is characterized by three
dimensionful constants: $\hL$ the size of the extra dimension
such that $z$ and $z + \hL$ are identified,
 $M$ the mass of the system measured at
infinity of ${\IR}^{d-2}$, and $G_N$ the Newton constant in $d$
dimensions. From these a single dimensionless parameter can be
constructed \be \label{mu_M}
 \mu=(G_N M)/\hL^{d-3}, \ee
while the $(d-1)$ dimensional effective Newton constant is given
by $G_{d-1}=G_N /\hL$.

The isometries of these solutions are $O(d-2) \times U(1) \times
(\IZ_2)^2$, where the $O(d-2)$ comes from the spherical symmetry
in ${\IR}^{d-2}$ which we assume and the $U(1)$ comes from time
independence.\footnote{In the Lorentzian solutions it is the
non-compact version of $U(1)$.} The discrete factors represents
reflections with respect to time and the $z$ coordinate $t \to
-t,\, z \to -z$. The most general metric with these isometries is
\be \label{general_metric}
 ds^2 = -e^{2\, \hA}\, dt^2 + ds^2_{(r,z)} + e^{2\, \hC}\, d\Omega^2, \ee
which is a general metric in the $(r,z)$ plane together with two
functions on the plane $\hA=\hA(r,z),\, \hC=\hC(r,z)$. The horizon is
a line determined by $e^{2\, \hA}=0$  and $d\Omega^2$ is the line
element for ${\IS}^{d-3}$.

There are several alternatives to fix the gauge and choose
coordinates in the $(r,z)$ plane, though for most of the current
theoretical analysis this does not matter. For the numerical
convenience we choose the ansatz  \be \label{ansatz}
 ds^2 = -A^2\, dt^2 + e^{2\, B}(dr^2 + dz^2) + r^2\, e^{2\, C}\,
 d\Omega^2 \label{conformal_ansatz} \ee
where the conformal gauge is chosen for the 2d metric and the new
variables $A,C$ are defined in terms of the variables in eq.
(\ref{general_metric}) \bea
 A &:=& e^{\hA} \non
C &:=& \hC -\log(r) \eea
 so that $A$ is regular at the horizon
($A|_{\mbox{hor}}=0$) and such that $C \to 0$ at infinity.

This ansatz still contains a gauge freedom for conformal
transformations which we finally fix by setting the domain of
definition for the three functions to be $\{(r,z): |z| \le L,
~r^2+z^2 \ge \rho_h^{~2} \}$, where we define $L:=\hL/2$ for
numerical convenience. It was shown in \cite{KudohTanakaNakamura}
that transforming to this domain is always possible by writing
down elliptic equations for the coordinate transformation. Thus
each domain is characterized by a dimensionless constant (the
conformal invariant) \be \label{defx}
 x := 2 \rho_h/\hL. \ee
The normalization was chosen so that $x=1$ is the maximum possible
value. Since the problem has only one dimensionless parameter all
quantities will be a function of $x$ (and in this sense it
replaces the dimensionless mass $\mu$).

We still need to state the boundary conditions. There are four
boundaries \begin{itemize}
 \item $r \to \infty$ infinity.

The metric is asymptotically flat \bea
 g_{\mu \nu} \sim \eta_{\mu \nu} &:=& \mbox{diag}(-1,1,\dots,1)  \Leftrightarrow \non
 \Leftrightarrow A &=& 1, ~B=C=0 \eea

 \item Horizon -- a regular horizon.

 \item $r=0$ axis -- regularity on the axis.

 \item $|z|=\hL/2$ -- periodicity (and reflection symmetry at $z=0$).
 \end{itemize}

By comparing with the Schwarzschild metric one can find the
relation $\rho=\rho(\rho_S)$ where $\rho_S$ is the Schwarzschild
coordinate (see appendix \ref{conformal_appendix}) \be {\rho \over
\rho_h} = \left[ ({\rho_S \over \rho_0})^{{d-3 \over 2}} +
\sqrt{({\rho_S \over \rho_0})^{d-3}-1 } \right] ^{{2 \over d-3}}
\ee where $\rho_S=\rho_0$ is the location of the horizon in
Schwarzschild coordinates. Asymptotic flatness chooses a natural
radial coordinate at infinity, hence we require that
asymptotically $\rho \sim \rho_S$, and we find the relation
between $\rho_h$ and $\rho_0$ \be \rho_h = \rho_0 / 2^{{2 \over
d-3}}. \label{rhoh_rho0}\ee

\section{Charges and the method of equivalent sources}
\label{equivalent_sources_section}

We start by analyzing the system at $r \to \infty$. In this limit
all the metric functions become independent of $z$ (since the mass
of the $z$-dependent modes is proportional to $1/\hL$ all $z$
dependence drops like $\exp(-r/\hL)$ as one gets away from the
black hole). Moreover, at infinity we are in the weak field limit,
namely the Newtonian limit.

It is possible to derive the properties of the asymptotics by
pretending that the black hole source is smeared over a much
larger area such that the gravitational field is everywhere weak,
while the energy-momentum tensor is non-vanishing (recall that
there are no sources in the sought-for black hole solutions). This
method is inspired by the analysis of
\cite{MyersPerry,MyersCompact}, and should be valid since far away
from the source it should not matter whether the source is a black
hole or a low density star. We term this approach ``a method of
equivalent sources''. After we derive the expression for the
charges in terms of the constants of asymptotics within this
method, we return in the next section and prove it rigorously
using the free energy and the first law.

One defines
 \bea g_{\mu \nu} &=& \eta_{\mu \nu} + h_{\mu \nu} \non
\bh_{\mu \nu} &:=& h_{\mu \nu}-\half\, h_\al^ \al \, \eta_{\mu \nu}
\label{defh_bh} \eea
where $\eta_{\mu \nu}$ is the flat space metric,
which is used for raising and lowering indices, and $h \ll 1$ is the
perturbation.  Choosing the harmonic (or ``Lorentz'') gauge
 \be
\del^\mu \, \bh_{\mu \nu} =0 \label{harmonic_gauge}, \ee
the linearized
field equations become \be \triangle \bh_{\mu \nu} = -16 \pi\, G_N\,
T_{\mu \nu}(x) \ee where $T_{\mu \nu}(x)$ is the stress energy tensor.
The solution may be written in a multipole
 expansion.\footnote{where the constant of proportionality is read as usual
 from $\triangle(1/r^{d-4})=- \Omega_{d-3}\, (d-4)\, \delta(\vec{x})$.}
The leading term at infinity comes from the monopoles (dipoles,
including angular momentum effects vanish here due to the
reflection symmetries, so the next correction will come from
quadruples), and in our geometry\footnote{In order to get the
solution in uncompactified
  spacetime replace $(d-4)\, \Omega_{d-3}\, \hL r^{d-4} \to (d-3)\,
  \Omega_{d-2}\, r^{d-3}$ in the following formulae.} it is
\be
\bh_{\mu \nu} = 2\, k_d {G_N \over (d-4)\, \hL\, r^{d-4}} T^Q_{\mu
  \nu} \ee
 where
\be T^Q_{\mu \nu} := \int d^{d-1}x T_ {\mu \nu} \ee
are ``total mass-energy charges'' at $t=\mbox{const}$, and
 \bea
k_d &:=& {8\, \pi \over \Omega_{d-3}} \label{kd} \\
\Omega_{d-1} &:=& d {\pi^{d/2} \over \Gamma(d/2 + 1)} = d {\pi^{d/2}
  \over (d/2)!} .  \eea
 $\Omega_{d-1}$ is the area of the unit
${\IS}^{d-1}$ sphere, and for odd $d$ we define $(d/2)! := \sqrt{\pi} {1
  \over 2} {3 \over 2}{5 \over 2} \dots {d \over 2}$. From now till
the end of this section we will use ``asymptotics adapted units''
such that $G_{d-1} = G_N/ \hL = 1$.

Finally one inverts the transformation (\ref{defh_bh}) to retrieve $h$
\bea
 h_{\mu \nu} &=& \bh_{\mu \nu}-{1 \over d-2} \, \bh_\al^\al \eta_{\mu \nu} = \non
 &=& {2\, k_d  \over  (d-4)\,  r^{d-4}} \left( T^Q_{\mu \nu} - {T^Q \over d-2} \eta_{\mu \nu} \right),
  \label{h_from_T} \eea
where $T^Q:= T^Q_{\al \bt}\, \eta^{\al \bt}$.

\subsection{Mass and tension}

The symmetries of the problem restrict the form of the energy-momentum
charges $T^Q_{\mu \nu}$: time reflection implies
$T^Q_{i0}=T^Q_{z0}=0$. Similarly, $z$ axis reflection implies
$T^Q_{zi}=0$. Spherical symmetry implies that $T^Q_{ij}=T^Q_{rr}\,
\delta_{ij}$. Here $i,j$ are any index other than $0\equiv t, z$.

It is important that not all three energy-momentum charges
$T^Q_{00},\, T^Q_{zz}\,\, T^Q_{rr}$ are independent, as a result
of GR's constraint equations. The relation we are about to derive
is a consequence of the constraint $G_{rr}=0$ (where $G$ is the
Einstein tensor), but here we show how to get the same result from
conservation of energy-momentum \be
 \del^\mu\, T_{\mu \nu} =0 \ee
 where we used $D_\mu =\del_\mu + O(h)$. In particular  \be
  \del^r\, T_{r r} =0 \Rightarrow T_{r r}(r) = \mbox{const} \label{Trr1} \ee
 but \be
  T_{r r} (r \gg \hL) =0 \Rightarrow T_{r r} (r)= 0 \ee
 and hence \be
 T^Q_{rr}=0. \label{Trr0} \ee

We denote \be
 m := T^Q_{00} ~, \ee
 the mass of the black hole.
$T_{zz}$ is pressure and since the conservation law tells us it is
constant along $z$: $\del_z\, T_{zz}=0$ it makes sense to define
\be
 \tau:=-T^Q_{zz}/\hL \ee
to be the tension along the periodic direction. This is the
tension measured by an asymptotic observer, which interprets it as
the tension of an imaginary string stretched along the compact
circle.
 The choice of sign reflects the fact that the
tension is always positive \cite{PositiveTraschen,PositiveSIT}.
Altogether we write the energy-momentum charges of the black hole
as
 \be T^Q_{\mu \nu} = \mbox{diag} (m,-\tau\, \hL,0,\dots,0) \label{T_charges} \ee

\subsection{From harmonic to conformal coordinates}

The most general ansatz for the asymptotic region in harmonic
coordinates (\ref{harmonic_gauge}) is \be ds^2 = -e^{-2\, A_H}\, dt^2 + e^{2\, B_H}\,
dz^2 + e^{2\, C_H}\, (dr_H^{~2} + r_H^{~2}\, d\Omega^2_{d-3} ) \label{harmonic_ansatz} \ee
where $A_H,B_H,C_H$ depend only on the harmonic radial coordinate
$r_H$, and the sign for $A_H$ was defined for later convenience.

Define the constants $a_H,\, b_H,\, c_H$ (with dimension
$\mbox{length}^{d-4}$) from the asymptotic form of the functions
\bea
 A_H &=& {a_H \over r^{d-4}} + O({1 \over r^{d-3}}) \non
 B_H &=& {b_H \over r^{d-4}} + O({1 \over r^{d-3}}) \non
 C_H &=& {c_H \over r^{d-4}} + O({1 \over r^{d-3}}). \label{def_harmonic_asymp} \eea

The harmonic gauge condition (\ref{harmonic_gauge}) sets some
relations between these constants. Actually, only $\del^r
\bh_{rr}$ is not identically zero and gives \be \fbox{$~~
 -a_H + b_H + (d-4)\, c_H=0 ~~$} ~~. \label{harmonic_constraint} \ee

Eqs. (\ref{h_from_T}, \ref{T_charges}) tell us that \be
  \left[ \begin{array}{c} a_H \\ b_H \\ c_H \\ \end{array} \right] =
  {k_d \over (d-4)\, (d-2)} \left[ \begin{array}{cc}
  d-3 & -1 \\
  1   & -(d-3) \\
  1   & 1 \\
   \end{array} \right] \,
 \left[ \begin{array}{c} m \\ \tau\, \hL \\ \end{array} \right] \label{charges_to_Hasymp} \ee
 Note that owing to this relation the constraint (\ref{harmonic_constraint}) is satisfied
automatically, which can be traced to the conservation of
energy-momentum (\ref{Trr1}).

Comparing with the conformal ansatz (\ref{conformal_ansatz}),
where this time the functions $A,\, B,\, C$ depend only on $r$,
one gets \bea
 A &=& e^{-A_H} \label{AAH} \\
 B &=& B_H \label{BBH} \\
 e^{B}\, dr &=& e^{C_H}\, dr_H \label{rH_to_c} \\
 C &=& C_H - \log(r/r_H) \label{CCH}
 \eea
The equation for $r$ from (\ref{rH_to_c}, \ref{BBH})

 \bea dr &=& e^{C_H-B_H}\, dr_H = \non
         &=& (1 + {c_H-b_H \over r_H^{~d-4}} +O({1 \over r_H^{~d-3}})
         ~)\, dr_H \eea
can be integrated to give
 \bea r=\left\{ \begin{array}{llr}
    &r_H +k_5\, \tau\, \hL\, \log({r_H \over r_{H0}}) + O({1 \over r_H}),
    &\mbox{for}\   d=5;  \label{rH_to_rd5} \\ \\

    &r_H + r_0- {k_d \ \tau\, \hL \over (d-4)\, (d-5)\, r_H^{~d-5}} + O({1 \over
   r_H^{~d-4}}),   &\mbox{for} \  d > 5 \label{rH_to_rdg5} ,
\end{array}\right.
\eea
where $r_0,\, r_{H0}$ are constants of integration which are
needed in the numerics in order to keep the location of the
horizon fixed. Note that the case $d=5$ is somewhat special and
needs to be treated separately.

We are now ready to find the constants in the conformal gauge
asymptotics. For $A,\, B$ the work is done already due to
(\ref{charges_to_Hasymp}-\ref{BBH}). We define $a,\, b$ by \bea
 1-A &=& {a \over r^{d-4}} + o({1 \over r^{d-4}}) \non
 B &=& {b \over r^{d-4}} + o({1 \over r^{d-4}}) \label{defab} \eea
and we get
 \be
  \left[ \begin{array}{c} a \\ b \\  \end{array} \right] =
  \left[ \begin{array}{c} a_H \\ b_H \\  \end{array} \right] =
  {k_d \over (d-4)\,(d-2)} \left[ \begin{array}{cc}
  d-3 & -1 \\
  1   & -(d-3) \\
   \end{array} \right] \,
 \left[ \begin{array}{c} m \\ \tau\, \hL \\ \end{array} \right] \label{charges_to_asymp} \ee
 which can be inverted to give \be \fbox{$~~
  \left[ \begin{array}{c} m \\ \tau\, \hL \\  \end{array} \right] =
  {1 \over k_d} \left[ \begin{array}{cc}
  d-3 & -1 \\
  1   & -(d-3) \\
   \end{array} \right] \,
 \left[ \begin{array}{c} a \\ b \\ \end{array} \right] ~~$} \label{asymp_to_charges} \ee
  (reminder: $k_d$ is defined in (\ref{kd})).

The last equation is one of our main results -- it tells us the
physical charges in terms of the numerical asymptotics.

We now turn to the function $C(r)$ which we treat separately for
$d=5$ and for $d>5$. For $d=5$ an expansion of (\ref{CCH}) using
the change of variables (\ref{rH_to_rd5}) gives \be
 C = {c_H + k_5\, \tau\, \hL \, \log(r_{H0)} \over r} - {k_5 \, \tau \, \hL \over r} \log(r)
 + O({ \log^2 r \over r^2}) \label{cd5} .\ee
  The leading term at infinity is $\log(r)/r$ and so we may define $c_5$ by \be
 C = c_5 {\log(r) \over r} + \dots ~~, \label{defc5} \ee
  and using (\ref{cd5}, \ref{asymp_to_charges}) we get \be \fbox{$~~
  c_5 = b_{H5}-c_{H5} = 2\, b -a .$} \label{5drelation}\ee
This constraint is the 5d manifestation of
(\ref{harmonic_constraint}). Note that the next term $O(1/r)$ is
hard to distinguish from the leading term and it contains an
arbitrary constant $r_{H0}$.

For $d>5$ (\ref{CCH}) yields \be
 C = {c_H \over r_H^{~d-4}} - \log(1 + {r_0 \over r_H} - {k_d\, \tau\, \hL \over (d-4)\, (d-5)\, r_H^{~d-4}}
 + O({1 \over r_H^{~d-5}}) ) \ee
 which together with the inversion of (\ref{rH_to_rdg5}) \be
 r_H= r-r_0 + {k_d\, \tau\, \hL \over (d-4)\, (d-5)\, r^{d-5}} + O({1 \over r^{d-4}}) \ee
tells us that the leading term in $C(r)$ is $-{r_0 \over r}$ while
the charge $\tau$ enters only at order $O({1 \over r^{d-4}})$ and
is mixed with $r_0$.

\subsection{Small black holes}

Myers \cite{MyersCompact} argues that for small black holes ($x
\ll 1$) one should take zero tension, $\tau=0$, namely the
equivalent source is dust. In this case we use
(\ref{charges_to_asymp}) to compute the asymptotics \be \fbox{$~~
 \left[ \begin{array}{c}
  a \\ b \\ \end{array} \right] \simeq {k_d \over (d-4)\,(d-2)}\, m\,  \left[ \begin{array}{c}
  d-3\\ 1\\ \end{array} \right] ~~$} ~~,  \label{small_asymp} \ee
 where the $\simeq$ sign denotes that the ratio of the two sides approaches $1$ as $x \to 0$.

Moreover, in this limit the mass is determined by the horizon size
\be \fbox{$~~
 {2\, k_{d+1} \over (d-2)\, \hL^{d-3}}\, G_N\, m\simeq ({\rho_0 \over \hL})^{d-3} = 4\,
  \left({x \over 2}\right)^{d-3} ~~$}
 \label{small_mass}~~.\ee
This can be seen by using the method of equivalent sources to
compute the mass of a Schwarzschild black hole \cite{MyersPerry}
and then noting that after compactification the mass does not
change in the weak gravity region since the Newtonian potential
satisfies Gauss' law.

\section{Charges, free energy and the first law}
\label{free_energy_section}

\subsection{Charges from free energy}

Let us derive the form of the first law (conservation of energy)
using rigorous free energy techniques. We will see that the
results agree with those of the previous section.

The (Gibbons-Hawking \cite{GibbonsHawking}) free energy action
integral is defined for the Euclidean continuation of static
metrics and is given by \be I = -\beta\, F = {1 \over 16\, \pi\,
G_N}\, \int dV_d\, R_d + {1 \over 8\,
  \pi\, G_N}\, \int_\del dV_{d-1}\, [K-K^{0}] \label{defF1}
 \ee where
$\beta=1/T$ is the inverse temperature (the period of the
Euclidean time), $F$ is the free energy, $R_d$ is the $d$
dimensional Ricci scalar, $dV_d,\, dV_{d-1}$ are the $d,\, (d-1)$
invariant volume elements, the second term is integrated over the
boundary of the manifold $\del$, $K$ is the trace of the second
fundamental form, while $K^{0}$ is the same quantity for a
reference geometry with the same boundary.

Consider the action to be a function of the boundary conditions
\be
 I=I(\beta,\hL) \ee
evaluated after extremization over metrics. In order to get the
first law we need an explicit expression for the action, find the
total differential $dI= (\del I/\del \beta)\, d\beta + (\del
I/\del \hL)\, d\hL$ and then use thermodynamic identities to
transform it into the standard form.

The action in terms of the harmonic fields (\ref{harmonic_ansatz})
reads (see appendix \ref{EvaluateI} for the derivation) \be
\fbox{$~~
 I = {\beta_0\, \hL_0  \over 2\, k_d\, G_N}
 \int dr_H\, r_H^{d-3}\, e^{-A_H+B_H+(d-4)\, C_H} ~ K_{ij}\,  X'_i\, X'_j
 ~~$} \label{harmonic_action} \ee
 where we use a short-hand notation $X_i=[A_H, ~B_H, ~C_H]$ and
 the matrix $K_{ij}$ is
 \be
 K= \left[ \begin{array}{ccc}
 0 &   -1 & -(d-3) \\
 -1 &   0 & (d-3) \\
 -(d-3) & (d-3) & (d-3)(d-4) \\ \end{array} \right] \ee
 Moreover, there is no boundary term.

Having the expression for the action we wish to compute its
differential. By a general theorem in mechanics
 \footnote{See for example \cite{LL}:  $\delta I = \left[{\del {\cal L}
\over \del \dot{q}}\, \delta q \right]_{t1}^{t2} + \int_{t1}^{t2}
({\del {\cal L} \over \del q}- {d \over dt} {\del {\cal L} \over
\del \dot{q} })\, \delta q\, dt$ and since the second term
vanishes as we are evaluating on solutions of the equations of
motion, only the first term contributes.} it is given by the
conjugate momentum at the boundary \be
 \del_{Xi} I = { \del {\cal L} \over \del_r X_i} ~~|_{r_H=R} \label{conjugate_momenta} \ee
 where ${\cal L}$ is the Lagrangian density.

Using the constants of asymptotics (\ref{def_harmonic_asymp}) we
may compute \be \left[ \begin{array}{c}
 \del_{A_H} \\ \del_{B_H} \\ \del_{C_H}
\end{array} \right] I = - { \beta_0\, \hL \over k_d \, G_N}(d-4)\, K\,
\left[
\begin{array}{c} a_H \\ b_H \\ c_H \end{array} \right]
 \label{delI_from_asymp1} \ee
 where (\ref{conjugate_momenta}) was evaluated at the $R \to
 \infty$ limit.

$\del_{C_H}$ is precisely proportional to the constraint
(\ref{harmonic_constraint}) (derived from $G_{rr}=0$) and thus
vanishes identically. This is to be expected since a shift in $C$
is equivalent to a shift of the boundary in $r_H$ without changing
the asymptotic sizes $\beta,\, \hL$, and since we are evaluating
$I$ in the $R \to \infty$ limit, the action should not change. Now
we may use the constraint (\ref{harmonic_constraint}) to eliminate
$c_H$ from the expressions (\ref{delI_from_asymp1}) and we may
drop the $H$ subscript from $a,b$ according to
(\ref{charges_to_asymp}) and get \be
 \left[ \begin{array}{c}
  \del_{A_H} \\ \del_{B_H}
 \end{array} \right] I
= { \beta_0\, \hL \over k_d \, G_N}\,
 \left[ \begin{array}{cc}
 d-3 & -1 \\
 -1 & d-3 \\
 \end{array} \right]
 \left[ \begin{array}{c}
   a \\ b
 \end{array} \right]
 \label{delI_from_asymp2} \ee

Finally the relations $d\beta=-\beta_0\, dA_H,
 ~d\hL = \hL_0\, dB_H$ valid asymptotically where $A_H=B_H=0$
 yield the differential of the action \be
 {k_d\, G_N \over \hL}\, dI = -((d-3)\, a-b)\, d\beta - {\beta \over \hL} (a-(d-3)\,
 b)\, d\hL \label{dI_from_asymp} \ee

Now let us compare with standard thermodynamics. In analogy with
gas thermodynamics where one has  \be
 dE = T\, dS - P\, dV \label{gas_first_law} \ee
the first law must take the following form \be \fbox{$~~
 dm = T\, dS + \tau\, d\hL ~~$}~~.\label{diff_first_law}\ee
 This can be considered the proper definition of the mass $m$ and the
tension $\tau$. We will now see that this definition coincides
with the definition in terms of asymptotics
(\ref{asymp_to_charges}) and hence all is consistent. The free
energy is related to $m$ by $F=m-T\, S$ and the action is
$I=-\beta\, F$. Putting the last two relations together one finds
\be
 dI = -m\, d\beta - \beta\, \tau \, d\hL \label{dI} \ee
Comparing with (\ref{dI_from_asymp}) we re-obtain the expressions
for the mass and tension (\ref{asymp_to_charges})
as anticipated above. 
Alternatively,  if one takes (\ref{asymp_to_charges}) to be the
definition of mass and tension, then (\ref{dI_from_asymp}) proves
the form of the first law (\ref{diff_first_law}) and suggest the
analogy with gas thermodynamics (\ref{gas_first_law}).

\subsection{Smarr's formula}

In this section we describe Smarr's formula, also known as the
integrated form of the first law, for the geometry under study. It
is a relation between thermodynamic quantities at the horizon with
those at infinity, relying on the generalized Stokes formula and
the validity of the equations of motion in the interior,  and as
such it estimates the ``overall numerical error'' in our numerical
implementation \cite{numerical}.

In general Smarr's law can be gotten from the first law
(\ref{diff_first_law})  after taking account of the scaling
dimensions. Performing a scaling transformation $\hL \to (1+\eps)
\hL$, recalling the relations $T=\kappa/(2\, \pi), ~S=A/(4\,
G_N)$, and that the area has dimension $d-2$ while the mass has
$d-3$, we get by we expanding (\ref{diff_first_law}) to first
order in $\eps$ \be \fbox{$~~
 (d-3)\, m = (d-2)\, {\kappa\, A \over 8\, \pi G_N} + \tau\, \hL ~~$}. \label{Smarr_formula} \ee

In order to put (\ref{Smarr_formula}) into a ``numerically adapted
form'', that is one that uses only the numerical quantities we use
(\ref{asymp_to_charges}) to get \be {1 \over 8\, \pi\, G_N} A\,
\kappa ={(d-4) \hL \over k_d\, G_N}\, a,
\label{smarr_for_numerics} \ee
 where on LHS we have horizon quantities and on the RHS asymptotic
 ones. Finally the LHS can be written also as $T\, S$.

\section{Dimensional reduction and the scalar charge}
\label{dim_reduction_section}

At infinity, since all $z$ dependence is lost, it is natural to
look at the system from a $(d-1)$ point of view. We will see that
this point of view offers a natural interpretation for the mass
formula (\ref{asymp_to_charges}) --  the mass can be read from the
metric in the usual way. In addition, this perspective suggests
the importance of the scalar charge to be discussed next.

The $d$ dimensional metric can be written as \be
 ds^2 = \tg^{(d-1)}_{ab}\, dx^a\, dx^b + e^{2\, \phi}\, dz^2 \ee
where $a,b$ run over the $d-1$ dimensions. Thus after reduction we
get a metric and a scalar field $\phi$ (for a general dimensional
reduction one gets a vector as well, but it vanishes in our static
 ansatz).

Computing the action after reduction we have (\ref{Rfibration})
\bea
 R^{(d)} &=& \tR^{(d-1)} + \dots \non
 \sqrt{-g^{(d)}} &=& e^{\phi}\, \sqrt{-\tg^{(d-1)}} \eea
where the $R$ is the Ricci scalar and $\dots$ could include other
terms such as $(\del \phi)^2,\, \triangle(\phi)$. One notices that
the gravitational action behaves as $e^{\phi}\,
\sqrt{-\tg^{(d-1)}}\, \tR^{(d-1)}$ and is not canonically
normalized. In order to mend that a Weyl rescaling is required \be
 \tg^{(d-1)} = e^{2\, w}\, g^{(d-1)}. \ee
The changes are (\ref{Rconformal}) \bea
 \tR^{(d-1)} &=& e^{-2\, w}\, \tR^{(d-1)} + \dots \non
 \sqrt{-\tg^{(d-1)}} &=& e^{(d-1)\, w}\, \sqrt{-g^{(d-1)}} \eea
where the $\dots$ denote this time terms with $(\del w)^2$ and
$\triangle(w)$. We see that in order to get the Einstein action we
should choose \be
 w = -{\phi \over d-3}. \ee

Let us look at $h^{(d-1)}_{00}$ \bea
 h^{(d-1)}_{00} &=& e^{-2\, w}\, \tg^{(d-1)}_{00}\, +1 \simeq \non
  &\simeq& (1 + {2\, b \over (d-3)\ r^{d-4}}) \, (-1+{2\, a \over
  r^{d-4}}) + 1 \simeq \non
 &\simeq& {2\, \over r^{d-4}}\, (a- {1 \over d-3}\, b) \eea
But this should equal $(\rho_0/r)^{d-4}$ hence \be
 2\, (a- {1 \over d-3}\, b) = (\rho_0)^{d-4}= {2\, k_d\, G_N \over
 (d-3)\, \hL}\, m \ee
 where the second equality comes from (\ref{small_mass}) and
altogether we reproduced (\ref{asymp_to_charges}).

Let us reinterpret the asymptotic constant $b$ in light of the
dimensional reduction. Since by definition $B_H = \phi$, $b$
describes now the fall-off of the scalar field $\phi$ \be
 \phi \simeq {b \over r^{d-4}} \label{defScalarCharge} \ee
Such a quantity is usually called {\it the scalar charge} (of
$\phi$). In \cite{ScalarCharge} it was shown that the scalar
charge appears in the first law as $(dm/d\phi^i_{\infty})=
G_{ij}\, \Sigma^j$, where $\Sigma^j$ are scalar charges defined in
analogy with $b$ in (\ref{defScalarCharge}), the indices $i,j$
allow for several scalar fields with a non-trivial metric read
from the kinetic term $G_{ij}(\phi)\, \del \phi^i\, \del \phi^j$.
Indeed after the Weyl rescaling $b$ replaces $\tau$ in the
expression for $dI$ (as in (\ref{dI}), while at the same time we
must use the transformed $\beta$.

We can now interpret the equation for $b$ (\ref{charges_to_asymp})
as telling us that the mass increases $b$ and tries to open up the
extra dimension, while the tension counteracts. For the uniform
string these two effects exactly cancel each other. Hence $b$ is a
useful order parameter to describe the departure from the uniform
string. Moreover, it has the advantage of being finite also for
the black hole and hence one can naturally put all the phases on
the same phase diagram as advocated in \cite{TopChange}. We notice
that $b$ is always positive as far as we know. It would be
interesting to understand/ refute this.

\section{At the horizon}
\label{at_horizon}

It turns out that all the measurable quantities which we define
reside either at infinity or at the horizon. After discussing the
former in the previous sections, we now turn to the latter. In the
vicinity of the horizon it is convenient to transform the $(r,z)$
plane into polar $(\rho,\chi)$ coordinates through $z=\rho\, \cos
(\chi),~ r=\rho\, \sin (\chi)$.

\subsection{Definition of measurables}

There are two thermodynamic quantities defined at the horizon \begin{itemize}
 \item The area, $A$.

 In the conformal gauge (\ref{conformal_ansatz}) it is given by \bea
 A &=& \int_0^\pi  \rho_h\, e^B\, d\chi \cdot \Omega_{d-3}\, (\rho_h\, e^C\, \sin(\chi))^{~d-3} = \non
 &=& \Omega_{d-3}\, \rho_h^{~d-2}\, \int_0^\pi   e^{B+(d-3)\, C}\, \sin(\chi)^{~d-3}\, d\chi \eea

 \item Surface gravity, $\kappa$.

Since the geometry is static we may use the definition of kappa by
taking the Euclidean continuation of the metric and finding the
period $\beta$ of Euclidean time such that there is no conical
singularity at the horizon. Then $\kappa = 2\, \pi/ \beta$. In
conformal coordinates it is given by \be
 \kappa = e^{-B}\, \del_\rho A \ee
The same expression is gotten from the equivalent definition
$\kappa^2 = -(1/2) (\del^\mu \xi^\nu)(\del_\nu \xi_\mu)$ where
$\xi^\mu$ is the Killing field and the expression is evaluated on
the horizon (see for example \cite{Wald}).

\end{itemize}

In addition we monitor two quantities that measure the geometry of
the horizon \begin{itemize}

\item The eccentricity, $\eps$.

  The spherical symmetry of a small black hole, $SO(d-1)$, is broken
  to $SO(d-2)$, as it grows. In the $\{r,z\}$ plane we define the
  eccentricity \footnote{This definition differs from the standard
    definition of an ellipse's eccentricity.  It is analogous to
    $a/b-1 = (1-e^2)^{-1/2} \simeq 1 + 0.5\, e^2$, where $e$ is the
    conventionally defined eccentricity.}
\bea \eps &:=& {A_{\perp}
    \over A_{\parallel}} -1 \non A_{\parallel} &:=& \Omega_{d-3}\,
  e^{(d-3)\, C}\, \rho_h^{~d-3} ,~~ {\rm at} \ z=0, \non A_{\perp} &:=& \int_0^\pi
  \rho_h\, e^B\, d\chi \cdot \Omega_{d-4}\, (e^C\, \rho_h\,
  \sin(\chi))^{~d-4} = \non &=& \Omega_{d-4}\, \rho_h^{~d-3}\,
  \int_0^\pi e^{B+(d-4)C}\, \sin(\chi)^{d-4}\, d\chi \eea

$A_{\parallel}, A_{\perp}$ are sections of the horizon:
~$A_{\parallel}$ is the area of the equatorial sphere at $z=0$,
while $A_{\perp}$ is the area of the axial (or ``polar'') sphere
at $\chi=0,\pi$.

\item The polar distance, $L_{\mbox{poles}}$.

This is the proper distance between the ``north'' and ``south'' poles along
the $r=0$ axis and it is given by \be
 L_{\mbox{poles}} := 2 \int_{\rho_h}^{\hL/2} dz\, e^{B} ,~~ {\rm at} \ r=0. \ee
\end{itemize}

\subsection{Predictions for small black holes}

Out of the two thermodynamic quantities $A,\, \kappa$ one can form
a dimensionless quantity %
\be
 A^{(\kappa)} := A\, \kappa^{d-2} \ee

For small black holes we may take the Schwarzschild metric as an
approximation, where $A=\Omega_{d-2}\, \rho_0^{~d-2},
~\kappa=(d-3)/(2\, \rho_0)$, and hence \be
 A^{(\kappa)} = A^{(\kappa)}_0 := \Omega_{d-2} \left( {d-3 \over 2} \right)^{d-2}. \ee

More generally we may wish to expand $A^{(\kappa)}=A^{(\kappa)}(x)$ in a Taylor series \be
 A^{(\kappa)}(x) = \sum_{j=0}^{\infty} A^{(\kappa)}_  j\, x^j \ee
Note that $A^{(\kappa)}_j$ depends both on $j$ and on the
dimension $d$ which is suppressed in this notation.

 It can be shown \cite{GorbonosKol} that the next non-vanishing term
 is $A^{(\kappa)}_{d-3}$ and it is given by
$$ A^{(\kappa)}_{d-3} =
 - A^{(\kappa)}_0\, (d-2)\, \zeta(d-3) \left({\rho_0 \over
     \hL}\right)^{d-3} \Rightarrow
$$
\be\fbox{$~~ A^{(\kappa)}_{d-3}\ /A^{(\kappa)}_0
   = - {4\, (d-2) \over 2^{d-3}} ~ \zeta(d-3) ~~$} ~, \label{Akappa} \ee
 where $\zeta$ is
 Riemann's zeta function, and we used eq. (\ref{rhoh_rho0}).

For the eccentricity the prediction is \cite{GorbonosKol} \be
 \eps = c_\eps\, x^{d-1} \ee
 while the constant of proportionality is currently computed only in 5d \be \fbox{$~~
 c_\eps (d=5) = {8 \over 3}\, \zeta(4) = {4\, \pi^4 \over 135} \simeq 2.89
 ~~$} \label{eccentricity}
     \ee

     All 5d constants stated in this subsection are confirmed
     numerically \cite{numerical}.
\section{Summary}
\label{summary}

We summarize our main results: \begin{itemize}

\item Asymptotics.

There are two constants of asymptotics. In the harmonic ansatz
(\ref{harmonic_ansatz}) there are 3 of them defined by
(\ref{def_harmonic_asymp}) and they satisfy the constraint
(\ref{harmonic_constraint}). In the conformal ansatz
(\ref{conformal_ansatz}) the constants of asymptotics, $a,\, b$,
are defined in (\ref{defab}), related to the harmonic constants in
(\ref{charges_to_asymp}), and expressed in terms of the system's
mass and tension (around the $z$ coordinate) in
(\ref{asymp_to_charges}). We stress the physical interpretation of
the tension.

5d is somewhat special and we can furthermore define the
asymptotic quantity $c_5$ in (\ref{defc5}), and express it in
terms of $a,b$ in (\ref{5drelation}).

For small black holes we have a prediction for all quantities in
terms of $x$ in (\ref{small_asymp}, \ref{small_mass}).

\item Thermodynamics.

We derive the (differential form of the) first law in
(\ref{diff_first_law}) and the integrated form in
(\ref{Smarr_formula}).

\item Horizon region.

We make predictions for small black holes for two dimensionless
quantities: $A\, \kappa^{d-2}$ (\ref{Akappa}) and the horizon
eccentricity (\ref{eccentricity}).

\end{itemize}

\vspace{0.5cm} \noindent {\bf Acknowledgements}

It is a pleasure to thank J. Bahcall for suggesting this
collaboration. BK thanks T. Wiseman and D. Gorbonos for
discussions and collaboration on related issues.

The authors are supported in part by the Israeli Science
Foundation.

\appendix
\section{Schwarzschild in conformal coordinates}
\label{conformal_appendix}

Let us transform the Schwarzschild metric into conformal
coordinates and find the relation between the radii $\rho_0$ and
$\rho_h$.
 The Schwarzschild solution is \bea
\label{Schw}
 ds^2 = -f\, dt^2 + f^{-1}\, d\rho_S^2 + \rho_S^2\, d\Omega_{d-2}^{~2} \non
 f = 1 - \left( {\rho_0 \over \rho_S} \right)^{d-3} \eea
 One considers a new radial variable $\rho=\rho(\rho_S)$ such that \bea
 f^{-1}\; d\rho_S^{~2} &=& e^{2\, B}\, d\rho^2 \non
 \rho_S^{~2} &=& e^{2\, B}\, \rho^2 \label{metric_transformation} \eea
in terms of which the metric becomes \bea
 ds^2 &=& -f\, dt^2 +  e^{2\, B} (d\rho^2 + \rho^2\, d\Omega_{d-2}) = \non
       &=& -f\, dt^2 +  e^{2\, B} [d\rho^2 + \rho^2\, (d\chi^2 + \sin^2(\chi)\, d\Omega_{d-3}) ] \eea

>From (\ref{metric_transformation}) one extracts the equation for
$\rho$ \be
 {d\rho \over \rho} = f^{-1/2}{d\rho_S \over \rho_S} \ee
The integral on the right can be solved by a change of variables
$\cosh(\beta)=(\rho_S/\rho_0)^{(d-3)/2}$ and we get \bea
 \rho/\rho_h &=& \exp [ {2 \over d-3} \mbox{arccosh}((\rho_S/\rho_0)^{(d-3)/2}) ] = \non
             &=& \left[ ({\rho_S \over \rho_0})^{{d-3 \over 2}} + \sqrt{({\rho_S \over \rho_0})^{d-3}-1 } \right]
              ^{{2 \over d-3}}. \eea

The metric in the new conformal (or isotropic) coordinates takes
the following simple form \bea
 ds^2 &=& -\left( {1 -\psi \over 1+\psi} \right)^2\, dt^2 + (1+\psi)^{4/(d-3)}\, [ d\rho^2+\rho^2 \, d\Omega_{d-2}^2 ] \non
 \psi &:=& \left( {\rho_h \over \rho} \right)^{d-3} ~. \eea

\section{Hawking-Horowitz mass}

Let us compute the mass directly from the Hawking-Horowitz formula
as an additional ``independent'' derivation.\footnote{This mass
definition coincides with the ADM mass when both are applicable.}
The Hawking-Horowitz formula \cite{HawkingHorowitz} for a static
vacuum solution is
 \be
m_{HH} =  -{1 \over 8\, \pi\ G_N} \del_{\hat{n}}[
\Sigma^{(d-2)}-^0\Sigma^{(d-2)}] \label{HHmass} \ee
where $\Sigma^{(d-2)}$ is the $(d-2)$ area of the boundary of a
$t=\mbox{const}$ slice, $^0\Sigma^{(d-2)}$ is the same quantity
for a reference geometry, and $\del_{\hat{n}}$ is a derivative
with respect to orthogonal motion of this boundary.

We evaluate (\ref{HHmass}) in the harmonic ansatz
(\ref{harmonic_ansatz}) for a boundary at $r_H=R$ and take the $R
\to \infty$ limit. The first term is
\bea
 \del_{\hat{n}}\, \Sigma^{(d-2)} &=& e^{-C_H}\, \del_{r_H}  [
 (\hL\, e^{B_H})\, \Omega_{d-3}\, (r_H\, e^{C_H})^{d-3} ]\mid_{r_H=R} = \non \non
 &=& \hL\, \Omega_{d-3} \, ~ e^{-C_H}\,\, \del_{r_H}  [ r_H^{~d-3} \, e^{B_H+(d-3)\,C_H} ] _{r_H=R} \simeq \non\non
 &\simeq&  \hL\, \Omega_{d-3} \, [ (1-{c_H \over R^{d-4}})\, (d-3)\, R^{d-4} + (b_H+(d-3)\, c_H) ] = \non\non
 &=&  \hL\, \Omega_{d-3} \, [(d-3)\, R^{d-4} + b_H]. \label{mHHbd_result}\eea

The reference $t=\mbox{const}$ geometry is flat: $ds^2= dz^2 +
dr^2 + r^2\, d\Omega_{d-3}$ with period $\hL_R=\hL\, e^{B_H(R)}$
and  $r=R\, e^{C_H(R)}$ to match with the boundary. The second
term is
\bea \del_{\hat{n}}\, ^0\Sigma^{(d-2)} &=& \del_r [ \hL_R\,
\Omega_{d-3}\, r^{d-3} ]|_{r=R\,
  e^{C_H(R)}}~  = \non\non
  &=& (\hL\, e^{B_H(R)})\, \Omega_{d-3}\, (d-3)\, R^{d-4}\, e^{(d-4)\, C_H(R)} =
\non\non
 &\simeq& \hL\, \Omega_{d-3} \,(d-3)\, [
R^{d-4}+(b_H+(d-4)\, c_H)] = \non\non
 &=& \hL\, \Omega_{d-3}\, (d-3)\, [ R^{d-4} + a_H]. \label{mHHref_result} \eea
where in the last equality we used the relation
(\ref{harmonic_constraint}).

Combining (\ref{mHHbd_result},\ref{mHHref_result}) according to
(\ref{HHmass}) we get   \be
 m_{HH} =   {\hL\, \over k_d\, G_N}\, ((d-3)\, a- b) = m  \ee
which agrees with (\ref{asymp_to_charges}), and we omitted the
superfluous $H$ subscripts (\ref{charges_to_asymp}).
\section{The action in the harmonic ansatz}
\label{EvaluateI}

Here we derive the form of the action integral (\ref{defF1})  the
action in terms of the harmonic fields (\ref{harmonic_ansatz}).
Actually the asymptotic region will suffice and there the fields
depend only on $r_H$. In order to evaluate Ricci's scalar it is
useful to have the formula for it in the presence of a general
fibration \bea
 ds^2 &=& ds^2_X + \sum_i\, e^{2\, F_i}\, ds^2_{Yi} \Rightarrow \non
 R &=& R_X + \sum_i\, [ e^{-2\, F_i}\, R_{Yi} - 2\, d_i\,
 \wttriangle (F_i) - d_i (\del F_i)^2] -\sum_{i,j}\, d_i\, d_j\,
 (\del F_i \cdot \del F_j) \label{Rfibration} \eea
 where the fibration fields depend only on the $x$ coordinates,
 $R_X, ~R_{Yi}$ are the Ricci scalars of the spaces $X,\, Y_i$,
$F_i=F_i(x)$, and the Laplacian ($\wttriangle$) and grad-squared
($\del \cdot \del$) are evaluated in the X space.

Reducing over $t,z$ one has \be
 R_d = R_{d-2} - 2\,(\del A_H)^2- 2\,(\del B_H)^2 + 2\,(\del
 A_H \cdot \del B_H) + 2\, \wttriangle A_H - 2\, \wttriangle B_H \label{Rd} \ee
 Note that this equation should be (and is) symmetric under the
exchange $B_H \leftrightarrow -A_H$.

Since the $d-2$ metric is conformal to flat space we may use the
expression for the Ricci scalar of a conformally transformed
metric (see for example \cite{Wald2}) \bea
 \widetilde{ds}^2  &=& e^{2\, w}\, ds^2 \Rightarrow \non
 \widetilde{R} &=& e^{-2\, w}\, [ R -2\, (\hat{d}-1)\, \triangle w-(\hat{d}-1)(\hat{d}-2)\, (\del w)^2]
 \label{Rconformal} \eea
 where $\hat{d}$ is the dimension of the space and the Laplacian
and grad-squared are evaluated in the non-tilded metric. In our
case $w=C_H, ~ \hat{d}=d-2$ and $R=0$, hence \be
 R_{d-2} = - e^{-2\, C_H} [-2\, (d-3)\, \triangle C_H -(d-3)(d-4)\, (\del
 C_H)^2] \label{Rd-2} \ee
where the flat space Laplacian can be written as $\triangle_{d-2}
= \del_{rH}^{~2} + (d-4)/r\, \del_{rH}$.

Putting (\ref{Rd},\ref{Rd-2}) together one has \bea
 d^dx \sqrt{g_d}\, R_d &=& -2\, dr_H\, r_H^{d-3}\, \beta_0\, \hL_0\, \Omega_{d-3}\,
 e^{-A_H+B_H+(d-4)\, C_H} \times \non
&\times& [ (\del^2_{rH}+ {d-3 \over r_H} \del_{rH})\, [-A_H + B_H
+(d-3) C_H] + \non
 &+&(d-4)\, C'_H\, (-A'_H + B'_H) + \non
 &+& {(d-3)(d-4) \over 2}\,C_H^{'~2} +  A_H^{'~2} + B_H^{'~2}
 -  A'_H \,B'_H ]
 \eea
where $\beta_0\, \hL_0$ are the periods of the $t$ and $z$
coordinates, the prime denotes a derivative with respect to $r_H$
and we used $\wttriangle=\del_{rH}^{~2} + (d-4)/r_H ~
\del_{rH}+(d-4)\, C'_H\, \del_{r_H}$.

Performing integration by parts we end with \be
 I = {\beta_0\, \hL_0  \over 2\, k_d\, G_N}
 \int dr_H\, r_H^{d-3}\, e^{-A_H+B_H+(d-4)\, C_H} ~ K_{ij}\,  X'_i\, X'_j
  \label{harmonic_action2} \ee
 where we use a short-hand notation $X_i=[A_H, ~B_H, ~C_H]$ and
 the matrix $K_{ij}$ is
 \be
 K= \left[ \begin{array}{ccc}
 0 &   -1 & -(d-3) \\
 -1 &   0 & (d-3) \\
 -(d-3) & (d-3) & (d-3)(d-4) \\ \end{array} \right] \ee
 Moreover, the original boundary term in (\ref{defF1}) is exactly
such that it cancels against the boundary term generated by the
integration by parts so that (\ref{harmonic_action2}) is the full
action.

\section{Another derivation of Smarr's formula}

Here we give a more direct derivation of the integrated form of
the first law (Smarr's formula). The free energy is evaluated in
terms of the asymptotics, and the derivation is completed by
combined it with the analogous expression for the mass, and the
general thermodynamic relation $F=E-T\, S$. The latter identity
relies on integration by parts to relate the horizon with
infinity, and thus saves us from performing this step explicitly
in our derivation.

The free energy is defined in (\ref{defF1}). Since we discuss
vacuum black hole solutions we have $R_d=0$, and so only the
second term contributes. It can be rewritten as \be
 I = -\beta\, F =  {1 \over 8\, \pi\, G_N}\, \del_{\hat{n}} [ \Sigma^{(d-1)}-^0\Sigma^{(d-1)} ] \label{defF2} \ee
where $\Sigma^{(d-1)}$ is the $(d-1)$ area of the boundary
($^0\Sigma^{(d-1)}$ is the same for the reference geometry), and
$\del_{\hat{n}}$ is a derivative with respect to orthogonal motion
of this boundary.

Let us compute the free energy in the harmonic ansatz
(\ref{harmonic_ansatz}) for a boundary at $r_H=R$ and take the $R
\to \infty$ limit. The first term is \bea
 \del_{\hat{n}}\, \Sigma^{(d-1)} &=& e^{-C_H}\, \del_{r_H}  [(\beta\, e^{-A_H})\, (\hL\, e^{B_H})\, \Omega_{d-3}\, (r_H\, e^{C_H})^{d-3} ]|_{r_H=R} = \non\non
 &=& \beta\, \hL\, \Omega_{d-3} \, ~ e^{-C_H}\,\, \del_{r_H}  [ r_H^{~d-3} \, e^{-A_H+B_H+(d-3)\,C_H} ] |_{r_H=R} = \non\non
 &\simeq&  \beta\, \hL\, \Omega_{d-3} \, [ (1-{c_H \over R^{d-4}})\, (d-3)\, R^{d-4} + (-a_H+b_H+(d-3)\, c_H) ] = \non\non
 &=& \beta\, \hL\, \Omega_{d-3} \, [(d-3)\, R^{d-4} + (b_H-a_H)]. \label{Fbd_result}\eea

The reference geometry is flat: $ds^2=-dt^2 + dz^2 + dr^2 + r^2\,
d\Omega_{d-3}$ with periods $\beta_R = \beta\, e^{-A_H(R)},\,
\hL_R=\hL\, e^{B_H(R)}$ and  $r=R\, e^{C_H(R)}$ to match with the
boundary. The second term is \bea
  \del_{\hat{n}}\, ^0\Sigma^{(d-1)} &=& \del_r~ [ \beta_R\, \hL_R\, \Omega_{d-3}\, r^{d-3} ]|_{(r=R\, e^{C_H(R)})} = \non\non
 &=& (\beta\, e^{-A_H(R)})\, (\hL\, e^{B_H(R)})\, \Omega_{d-3}\, (d-3)\, R^{d-4}\, e^{(d-4)\, C_H(R)} = \non\non
 &\simeq&  \beta\, \hL\, \Omega_{d-3} \,(d-3)\, [ R^{d-4}+(-a_H+b_H+(d-4)\, c_H)] = \non\non
 &=& \beta\, \hL\, \Omega_{d-3}\, (d-3)\,  R^{d-4}. \label{Fref_result} \eea
where in the last equality we used the constraint
(\ref{harmonic_constraint}).

Combining (\ref{Fbd_result},\ref{Fref_result}) according to
(\ref{defF2}) we get \be
 -\beta\, F = {\beta\, \hL\, \over k_d\, G_N}\, (b_H- a_H)~, \ee
 namely \be \fbox{$~~
 F =   {\hL\, \over k_d\, G_N}\, (a- b) ~~$}~~ \ee
  where we omitted the superfluous $H$ subscripts (\ref{charges_to_asymp}).

Now we may use the relation \be
 F = m - T\, S \ee
 where $T=\kappa/(2\, \pi)$ is the temperature, and $S=A/(4\, G_N)$ is the entropy.
This relation holds for black hole thermodynamics and is proven
using integration by parts. Substituting $m$ from
(\ref{asymp_to_charges}) we get \be
 T\, S = {(d-4) \hL \over k_d\, G_N}\, a.
\ee
which coincides with the ``numerically adapted form'' of Smarr's
formula (\ref{smarr_for_numerics}).



\begin{thebibliography}{99}

\bibitem{Tangherlini}
F.~R.~Tangherlini, Nuovo Cimento {\bf 27}, 636 (1963).

\bibitem{GL1}
R.~Gregory and R.~Laflamme, ``Black Strings And P-Branes Are
Unstable,'' Phys.\ Rev.\ Lett.\  {\bf 70}, 2837 (1993)
[arXiv:hep-th/9301052].

\bibitem{GL2}
R.~Gregory and R.~Laflamme, ``The Instability of charged black
strings and p-branes,'' Nucl.\ Phys.\ B {\bf 428}, 399 (1994)
[arXiv:hep-th/9404071].

\bibitem{TopChange}
B.~Kol,
``Topology change in general relativity and the black-hole black-string transition,''
arXiv:hep-th/0206220.

\bibitem{Gubser}
S.~S.~Gubser, ``On non-uniform black branes,''
arXiv:hep-th/0110193.

\bibitem{Wiseman1}
T.~Wiseman,
``Static axisymmetric vacuum solutions and non-uniform black strings,''
Class.\ Quant.\ Grav.\  {\bf 20}, 1137 (2003)
[arXiv:hep-th/0209051].

\bibitem{numerical}
E.~Sorkin, B.~Kol, and T.~Piran, ``Caged black holes: black holes
in compactified spacetimes II -- 5d numerical implementation,'' to
appear.

\bibitem{PiranSorkin}
E.~Sorkin and T.~Piran, ``Initial data for black holes and black
strings in 5d,'' Phys.\ Rev.\ Lett.\  {\bf 90}, 171301 (2003)
[arXiv:hep-th/0211210].

\bibitem{HorowitzMaeda1}
G.~T.~Horowitz and K.~Maeda, ``Fate of the black string
instability,'' Phys.\ Rev.\ Lett.\  {\bf 87}, 131301 (2001)
[arXiv:hep-th/0105111].

\bibitem{CLOPPV}
M.~W.~Choptuik, L.~Lehner, I.~Olabarrieta, R.~Petryk, F.~Pretorius
and H.~Villegas, ``Towards the final fate of an unstable black
string,'' Phys.\ Rev.\ D {\bf 68}, 044001 (2003)
[arXiv:gr-qc/0304085].

\bibitem{deSmet}
P.~J.~Smet, ``Black holes on cylinders are not algebraically
special,'' arXiv:hep-th/0206106.


\bibitem{HO1}
T.~Harmark and N.~A.~Obers, ``Black holes on cylinders,'' JHEP
{\bf 0205}, 032 (2002) [arXiv:hep-th/0204047].

\bibitem{uniqueness}
B.~Kol, ``Speculative generalization of black hole uniqueness to
higher  dimensions,'' arXiv:hep-th/0208056.

\bibitem{explosive}
B.~Kol, ``Explosive black hole fission and fusion in large extra
dimensions,'' arXiv:hep-ph/0207037.

\bibitem{Wiseman2}
T.~Wiseman,
``From black strings to black holes,''
Class.\ Quant.\ Grav.\  {\bf 20}, 1177 (2003)
[arXiv:hep-th/0211028].

\bibitem{KolWiseman}
B.~Kol and T.~Wiseman,
``Evidence that highly non-uniform black strings have a conical waist,''
Class.\ Quant.\ Grav.\  {\bf 20}, 3493 (2003)
[arXiv:hep-th/0304070].

\bibitem{KudohTanakaNakamura}
H.~Kudoh, T.~Tanaka and T.~Nakamura,
``Small localized black holes in braneworld: Formulation and numerical  method,''
Phys.\ Rev.\ D {\bf 68}, 024035 (2003)
[arXiv:gr-qc/0301089].

\bibitem{EmparanMyers}
R.~Emparan and R.~C.~Myers,
``Instability of ultra-spinning black holes,''
JHEP {\bf 0309}, 025 (2003)
[arXiv:hep-th/0308056].

\bibitem{SusskindGW}
L.~Susskind, ``Matrix theory black holes and the Gross Witten
transition,'' [arXiv:hep-th/9805115].

\bibitem{HorowitzMaeda2}
G.~T.~Horowitz and K.~Maeda, ``Inhomogeneous near-extremal black
branes,'' arXiv:hep-th/0201241.

\bibitem{Lehner_review}
L.~Lehner, ``Numerical relativity: A review,'' Class.\ Quant.\
Grav.\  {\bf 18}, R25 (2001) [arXiv:gr-qc/0106072].

\bibitem{Reall}
H.~S.~Reall, ``Classical and thermodynamic stability of black
branes,'' Phys.\ Rev.\ D {\bf 64}, 044005 (2001)
[arXiv:hep-th/0104071].

\bibitem{Hubeny:2002xn}
V.~E.~Hubeny and M.~Rangamani, ``Unstable horizons,'' JHEP {\bf
0205}, 027 (2002) [arXiv:hep-th/0202189].

\bibitem{Casadio:2000py}
R.~Casadio and B.~Harms, ``Black hole evaporation and large extra
dimensions,'' Phys.\ Lett.\ B {\bf 487}, 209 (2000)
[arXiv:hep-th/0004004].

\bibitem{FrolovSquared}
A.~V.~Frolov and V.~P.~Frolov,
``Black holes in a compactified spacetime,''
Phys.\ Rev.\ D {\bf 67}, 124025 (2003)
[arXiv:hep-th/0302085].


\bibitem{TamakiKannoSoda}
T.~Tamaki, S.~Kanno and J.~Soda,
``Radionic non-uniform black strings,''
arXiv:hep-th/0307278.


\bibitem{Argyres:1998qn}
P.~C.~Argyres, S.~Dimopoulos and J.~March-Russell, ``Black holes
and sub-millimeter dimensions,'' Phys.\ Lett.\ B {\bf 441}, 96
(1998) [arXiv:hep-th/9808138].

\bibitem{Emparan:2000rs}
R.~Emparan, G.~T.~Horowitz and R.~C.~Myers, ``Black holes radiate
mainly on the brane,'' Phys.\ Rev.\ Lett.\  {\bf 85}, 499 (2000)
[arXiv:hep-th/0003118].

\bibitem{Giddings:2001bu}
S.~B.~Giddings and S.~Thomas, ``High energy colliders as black
hole factories: The end of short  distance physics,'' Phys.\
Rev.\ D {\bf 65}, 056010 (2002) [arXiv:hep-ph/0106219].


\bibitem{MyersPerry}
R.~C.~Myers and M.~J.~Perry, ``Black Holes In Higher Dimensional
Space-Times,'' Annals Phys.\  {\bf 172}, 304 (1986).

\bibitem{MyersCompact}
R.~C.~Myers, ``Higher Dimensional Black Holes In Compactified
Space-Times,'' Phys.\ Rev.\ D {\bf 35}, 455 (1987).


\bibitem{Townsend:2001rg}
P.~K.~Townsend and M.~Zamaklar, ``The first law of black brane
mechanics,'' Class.\ Quant.\ Grav.\  {\bf 18}, 5269 (2001)
[arXiv:hep-th/0107228].


\bibitem{PositiveTraschen}
J.~Traschen, ``A positivity theorem for gravitational tension in
brane spacetimes,'' arXiv:hep-th/0308173.

\bibitem{PositiveSIT}
T.~Shiromizu, D.~Ida and S.~Tomizawa, ``Kinematical bound in
asymptotically translationally invariant spacetimes,''
arXiv:gr-qc/0309061.


\bibitem{HO2}
T.~Harmark and N.~A.~Obers,
``New Phase Diagram for Black Holes and Strings on Cylinders,''
arXiv:hep-th/0309116.

\bibitem{HO3}
T.~Harmark and N.~A.~Obers, ``Phase Structure of Black Holes and
Strings on Cylinders,'' arXiv:hep-th/0309230.

\bibitem{GibbonsHawking}
G.~W.~Gibbons and S.~W.~Hawking, ``Action Integrals And Partition
Functions In Quantum Gravity,'' Phys.\ Rev.\ D {\bf 15}, 2752
(1977).

\bibitem{Wald2}
R.~M.~Wald, ``General Relativity,'' appendix D,
 {\it The University of Chicago Press}, 1984.

\bibitem{LL}
L.~D.~Landau and E.~M.~Lifshitz,
 ``Mechanics'' par. 43,
  {\it Pergamon Press}, 1976.

\bibitem{ScalarCharge}
G.~W.~Gibbons, R.~Kallosh and B.~Kol, ``Moduli, scalar charges,
and the first law of black hole thermodynamics,'' Phys.\ Rev.\
Lett.\  {\bf 77}, 4992 (1996) [arXiv:hep-th/9607108].

\bibitem{Wald}
R.~M.~Wald, ``General Relativity,'' eq. (12.5.14),
 {\it The University of Chicago Press}, 1984

\bibitem{GorbonosKol}
D.~Gorbonos and B.~Kol,
to be published.

\bibitem{HawkingHorowitz}
S.~W.~Hawking and G.~T.~Horowitz, ``The Gravitational Hamiltonian,
action, entropy and surface terms,'' Class.\ Quant.\ Grav.\  {\bf
13}, 1487 (1996) [arXiv:gr-qc/9501014].




\end{thebibliography}
\end{document}